\documentclass[amssymb,twocolumn,aps]{revtex4-2}
\usepackage{graphicx}
\usepackage{enumerate}
\usepackage{amssymb}
\usepackage{amsmath}
\usepackage{amsfonts}
\usepackage{color}

\begin{document}

\title{Asymmetric scattering of mirror symmetric radiation from nanostructures coupled to chiral films}

\author{A. Ciattoni$^1$}
\email{alessandro.ciattoni@spin.cnr.it}
\affiliation{$^1$CNR-SPIN, c/o Dip.to di Scienze Fisiche e Chimiche, Via Vetoio, 67100 Coppito (L'Aquila), Italy}

\date{\today}

\begin{abstract}
The interaction of radiation with chiral molecular films is not macroscopically invariant under mirror reflections and, accordingly, chiroptical effects exist which affect the spatial symmetry of the radiation profile and which nearly exclusively show up in the near-field due to the large molecule-wavelength size mismatch. Here we prove that the scattering of a mirror symmetric pair of plane waves by a nanowire lying on a chiral nanofilm is not mirror symmetric with an angular dissymmetry factor that can be as large as some tenths. Due to evanescent coupling, the nanowire efficiently experiences molecular chirality which produces a spatially asymmetric near-field so that the self-consistent unbalanced excitation of nanowire photonic modes with opposite angular momenta yields asymmetric far-field interference. In addition to enriching the physical understaning of mirror symmetry breaking in chiral nanophotonics, our results could suggest ultra-efficient schemes for enantiomeric discrimination which is essential in biological chemistry and pharmacology.
\end{abstract}

\maketitle

The investigation of molecular chirality is of crucial interest in science since it both involves fundamental physical concepts, as mirror symmetry breaking, and it is essential in biology and life sciences, for many organic molecules are chiral \cite{Wagni}. Due to the lack of molecular symmetry, both electric and magnetic dipoles assist the interaction of radiation with a chiral molecule and accordingly molecular polarizability is sensitive to both electric and magnetic fields \cite{Berov}. Since these two fields have different behavior under reflections, mirror symmetry breaking characterizes any chiroptical effect \cite{Barro}, the main example being optical activity where chiral molecules differently interact with left and right circularly polarized photons; this enables chiro-sensitive molecular detection through circular dichroism (the differential absorption of left and right circularly polarized light). However, since the magnetic dipole contribution is much smaller that the electric one (due to the mismatch between molecular size and radiation wavelength), differential chiroptical signals are usually very weak and a number of enhancement strategies have been proposed \cite{Colli,Munnn,Muuuu} as, for example, plasmon enhancement circular dichroism \cite{Govor,Abdul,Liuuu,Neste} where the scale mismatch is reduced by resorting to evanescent waves. Amplification of the differential circular dichroism signal has also been achieved with superchiral fields \cite{Tang1,Tang2,Hend1}, characterized by locally large optical chirality \cite{Lipki,Vazqu}, both in plasmonic \cite{Hend2,Schaf,Staub} and dielectric setups \cite{Moham,Garci}. Large differential responses of chiral molecules have similarly been identified in photoelectron circular dichroism \cite{Janss}, photoexcitation circular dichroism \cite{Beaul} and signal-reversing cavity ringdown polarimetry \cite{Sofik}.

In addition to the microscopic asymmetry of its molecules, a homogeneous chiral molecular film also displays macroscopic asymmetry since its mirror image in a normal mirror is an identical film but with opposite chirality (i.e. filled by the opposite enantiomeric molecules). As a consequence, besides sensitivity to polarization handedness, the chiroptical response of a chiral molecular film is also sensitive to the spatial symmetry of the radiation profile. The theoretical spatial symmetry analysis of the interaction between radiation and a chiral-molecular film has been performed in Ref.\cite{Ciat1} where, in particular, it has been shown that a chiral film probed by a mirror symmetric field of a nanoemitter produces an asymmetric near-field (mirror optical activity). An analogous microwave effect displaying marked spatial asymmetry has been experimentally observed in chiral metamaterials \cite{Hisam}.

In this Letter we show that marked asymmetry can be observed in the angular distribution of the radiation scattered by a nanowire lying on a chiral molecular nanofilm once illuminated by a mirror symmetric pair of plane waves. Basically the plane waves trigger the nanowire-film coupling whose evanescent character enhances the chiral respose of the film and produces a spatially asymmetric near-field by mirror optical activity. As a consequence, photonic nanowire modes with opposite angular momentum are differently excited and their interference in the far field produces the asymmetry of the scattered radiation pattern. Remarkably, even though the scattering asymmetry in the considered symmetric setup stems from a near-field chiroptical mechanism, both excitation and observation of the effect only rely on propagating radiation fields, as distinct from Refs.\cite{Ciat1,Hisam}. Besides, if the incident plane waves direction is close to trigger total internal reflection, the powers scattered by the nanowire into its lateral right and left sides can become very different with dissymmetry factor of about some tenths, a giant asymmetry considering that it entirely results from molecular chirality.

\begin{figure*} \label{Fig1}
\center
\includegraphics[width=1\textwidth]{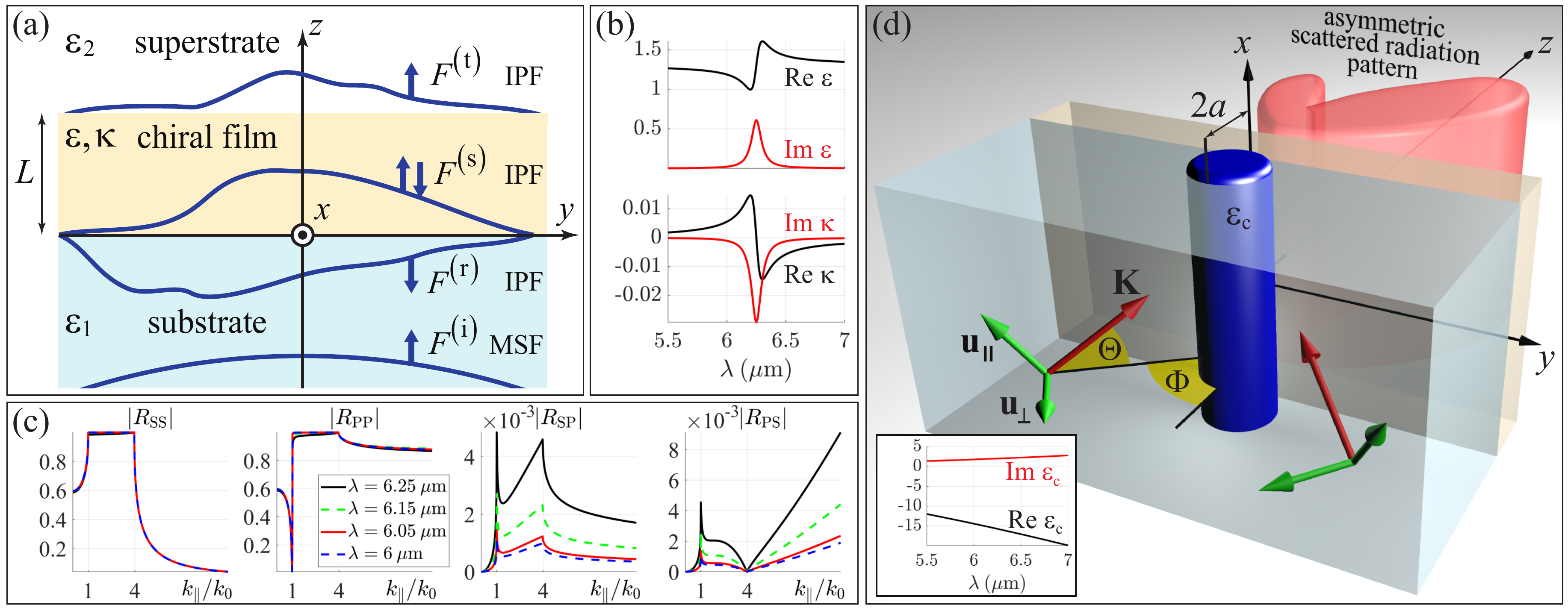}
\caption{(a) A mirror symmetric field (MSF) $F^{(\rm i)}$ illuminating a chiral film is reflected and transmitted into idenfinite parity fields $F^{(\rm r)}$, $F^{(\rm t)}$ (IPFs) as a consequence of the parity indefiniteness of the slab field $F^{(\rm s)}$ (mirror optical activity). (b) Permittivity and chirality paramenter of the chiral film  used for numerical evaluations, displaying molecular resonance at  $\lambda_0 = 6.25 \: \mu{\rm m}$. (c) Reflection coefficients of a $L = 50 \: {\rm nm}$ thick chiral film sandwitched between germanium ($\varepsilon_1 = 15.731$) and vacuum ($\varepsilon_2 = 1$) as functions of the (normalized) transverse photon momentum, at three wavelengths close to molecular resonance. (d) Setup of panel (a) equipped with a transparent conductor nanowire of radius $a= 30 \: {\rm nm}$ and permittivity $\varepsilon_c$ (reported in the inset) lying on the chiral film and embedded into the substrate. A mirror symmetric pair of plane waves (wave vector ${\bf K}$ and polarization basis ${\bf u}_\parallel$,${\bf u}_\bot$) is launched from the substrate and produces an asymmetric scattered radiation pattern in the superstrate.
}
\end{figure*}

We start by reviewing electromagnetic parity indefinitess  (and related mirror optical activity) in homogeneous chiral media where a monochromatic field ($e^{-i\omega t}$) is described by Maxwell's equations

\begin{equation} \label{MaxwellTEXT}
\def\arraystretch{1}
\left( {\begin{array}{@{\mkern0mu} c c @{\mkern0mu}}
   {\nabla  \times } & 0  \\
   0 & {\nabla  \times }  \\
\end{array}} \right)F = ik_0 \left( {\begin{array}{@{\mkern0mu} c c @{\mkern0mu}}
   {i\kappa } & 1  \\
   { - \varepsilon } & {i\kappa }  \\
\end{array}} \right)F,
\end{equation}
where $k_0 = \omega/c$, $\varepsilon$ and $\kappa$ are the permittivity and chiral (Pasteur) parameter and the electric and magnetic amplitudes $\bf E, H$ have been arranged as $F = ( {\begin{array}{@{\mkern0mu} l r @{\mkern0mu}} {\bf{E}} & {Z_0 {\bf{H}}} \end{array}} )^T$ ($Z_0 = \sqrt{\mu_0/\epsilon_0}$ is the vacuum impedance). Mirror symmetry breaking of Eq.(\ref{MaxwellTEXT}) is easily proved by noting that, for an arbitrary reflection through a plane (${\bf r'} = \mathcal{R} {\bf r}$) described by the dyadic $\mathcal{R}$, the mirror image of the field
\begin{equation}
\def\arraystretch{1}
F'\left( {\bf{r}} \right) = \left( {\begin{array}{@{\mkern0mu} c c @{\mkern0mu}}
   \mathcal{R} & 0  \\
   0 & { - \mathcal{R}}  \\
\end{array}} \right)F\left( {\mathcal{R}{\bf{r}}} \right),
\end{equation}
does not satisfies Eq.(\ref{MaxwellTEXT}) but and identical equation with $-\kappa$ replacing $\kappa$. This implies that the situations $F=F'$ (mirror symmetric field) and $F=-F'$ (mirror antisymmetric field) are ruled out or, in other words, that $F$ is an indefinite parity field (IPF, see Sec.S1 of Supplementary Material). Consider now the setup of Fig.1(a) where a chiral film of thickenss $L$ is sandwitched between two achiral media (substrate and superstrate of permittivities $\varepsilon_1$ and $\varepsilon_2$) and an incident field $F^{(\rm i)}$ is launched onto the film thus producing reflected and trasmitted fields $F^{(\rm r)}$, $F^{(\rm t)}$ together with the slab field $F^{(\rm s)}$. Focusing hereafter on reflections through the normal $xz$ plane, i.e. 
\begin{equation}
{\mathcal{R}} = {{\bf{e}}_x {\bf{e}}_x - {\bf{e}}_y {\bf{e}}_y  + {\bf{e}}_z {\bf{e}}_z },
\end{equation}
the field $F^{(\rm s)}$ is always an IPF by parity indefiniteness so that $F^{(\rm r)}$ and $F^{(\rm t)}$ are necessarily IPFs as well, even when the incident field $F^{(\rm i)}$ is a mirror symmetric field (MSF); the generation of reflected and transmitted IPFs by a chiral film once illuminated by a MSF is referred to as mirror optical activity (MOA), a chiroptical effect analyzed in Ref.\cite{Ciat1}. The planar geometry of the setup allows  analytical description in momentum space where the fields $F^{(\rm j)}$ ($\rm j=i,r,t$) are represented by their $\rm S$ (TE) and $\rm P$ (TM) complex amplitudes $U^{(\rm j)}_{\rm S}({\bf k}_\parallel)$, $U^{(\rm j)}_{\rm P}({\bf k}_\parallel)$ where ${\bf k}_\parallel = k_x {\bf e}_x + k_y {\bf e}_y$ is the parallel wavevector; reflection from the slab is for example descibed by the relation 
\begin{equation} \label{reflection}
\left( {\begin{array}{@{\mkern0mu} c c @{\mkern0mu}}
   {U_{\rm S}^{\left( \rm r \right)} }  \\
   {U_{\rm P}^{\left( \rm r \right)} }  \\
\end{array}} \right) = \left( {\begin{array}{@{\mkern0mu} c c @{\mkern0mu}}
   {R_{\rm SS} } & {nR_{\rm SP} }  \\
   {nR_{\rm PS} } & {R_{\rm PP} }  \\
\end{array}} \right)\left( {\begin{array}{@{\mkern0mu} c c @{\mkern0mu}}
   {U_{\rm S}^{\left( \rm i \right)} }  \\
   {U_{\rm P}^{\left( \rm i \right)} }  \\
\end{array}} \right)
\end{equation}
where $n=\sqrt{\varepsilon \kappa^2}/\kappa$ and the reflection coefficients $R_{\rm IJ}$ depend on $k_\parallel = |{\bf k}_\parallel|$ and $\kappa^2$. The mixing coefficients $R_{\rm SP}$ and $R_{\rm PS}$ in Eq.(\ref{reflection}) account for the coupling, produced by film chirality, between the $\rm S$ and $\rm P$ parts of the field and they are crucially responsible for MOA since, if $F^{\rm (\rm i)}$ is a MSF, the antisymmetric part of the reflected field, $[F^{\rm (r)} - {F^{\rm (r)}}' ]/2$, does not vanish since its complex amplitudes are $( {\begin{array}{@{\mkern0mu} l r @{\mkern0mu}} {nR_{\rm SP} U_{\rm P}^{\left(\rm  i \right)} } & {nR_{\rm PS} U_{\rm S}^{\left(\rm  i \right)} } \end{array}} )^T$ (see Sec.S2 of Supplementary Material). 

In this Letter we consider, for numerical evalutations, a $L = 50 \: {\rm nm}$ thick chiral film deposited on germanium ($\varepsilon_1 = 15.731$) in vacuum ($\varepsilon_2 = 1$) and consisting of a dielectric matrix hosting dispersed molecules of $\alpha$-alanine, characterized by marked chiral response around the molecular vibrational resonance at $\lambda_0 = 6.25 \: \mu{\rm m}$ (Sec.S1 of Supplementary Material) \cite{Tulio,Jahni}, as shown in Fig.1(b). In Fig.1(c) we plot the moduli of the reflection coefficients as functions of the normalized parallel photon momentum (for different wavelengths close to molecular resonance) markatedly signaling the special role played by germanium and vacuum wavenumbers $k_1 \simeq 4 k_0$ and $k_2 = k_0$, respectively. The diagonal coefficients  $R_{\rm SS}$ and $R_{\rm PP}$ are almost unaffected by chirality whereas the mixing coefficients $R_{\rm SP}$ and $R_{\rm PS}$ are maximum at molecular resonance (solid black lines) and very small far from it (see Sec.S2 of Supplementary Material), thus confirming their chiroptical nature. Moreover such mixing coefficients are not negligible only in the large photon momentum regime $k_\parallel > k_0$ so that MOA is effectively a nanophotonic effect triggered by fields $F^{\rm (i)}$ with deep subwelength spatial features (as, e.g., the field generated by a nanosource).

\begin{figure} \label{Fig2}
\center
\includegraphics[width=0.5\textwidth]{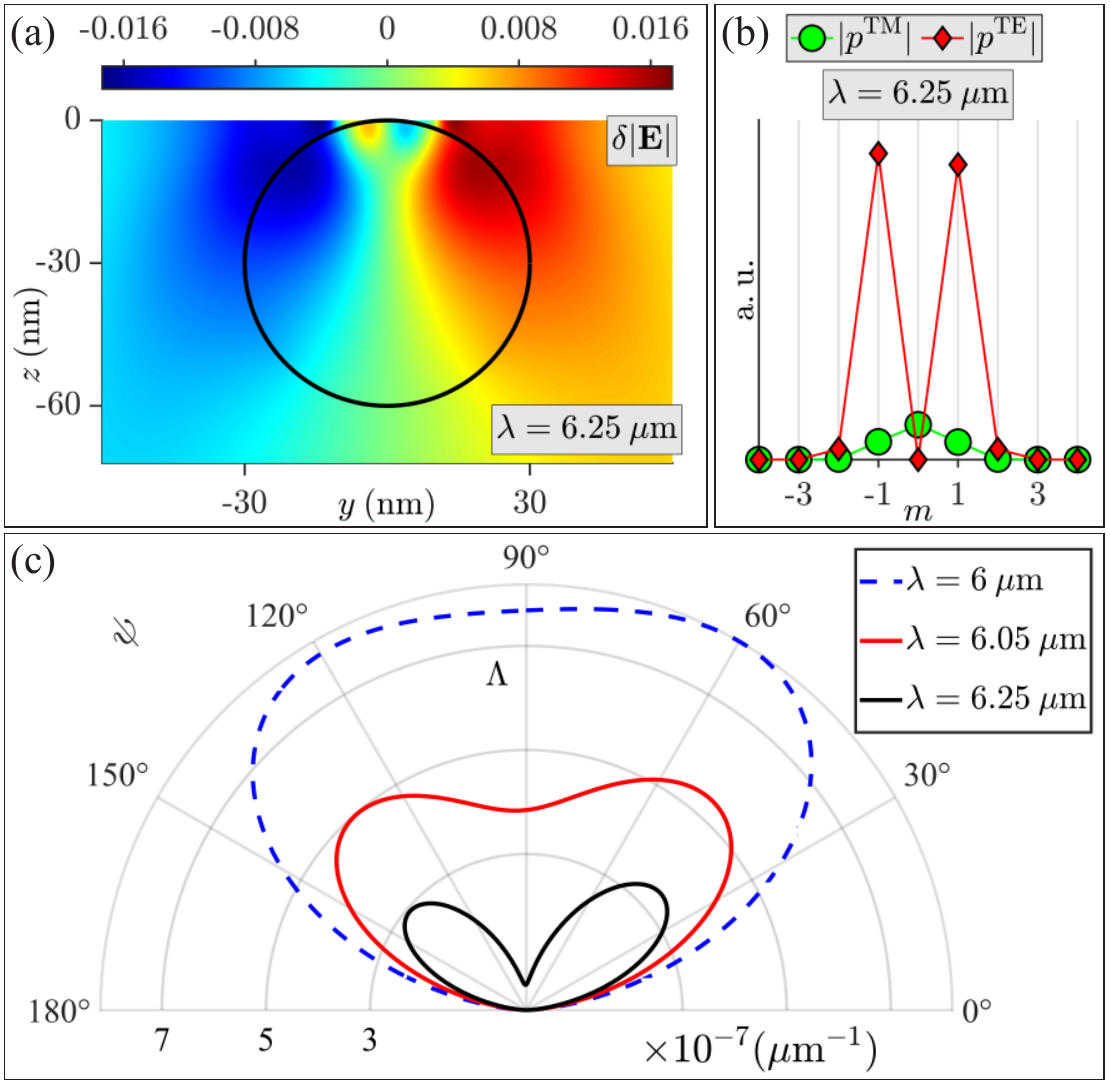}
\caption{Nanowire excitation and radiation scattering for the incident plane waves state $\Theta = 13.48^\circ$, $\Phi = 11.39^\circ$, $e_\parallel = 1/\sqrt{2}$ and $e_\perp = e^{i \pi/3} /\sqrt{2}$. (a) Spatial dissymmetry factor of the electric field $\delta |{\bf E}|$ (see Eq.(\ref{sdf})) in the substrate ($z<0$) at molecular resonance. (b) Moduli of the nanowire modes coefficients ${p^{\rm TM} \left( m \right)}$ and ${p^{\rm TE} \left( m \right)}$ at molecular resonance. (c) Polar plot of the scattering differential cross length $\Lambda(\psi)$ at different wavelengths. 
}
\end{figure}

Now the questions naturally arise as to whether there is a way to triggering MOA by means of propagating fields (not comprising evanescent waves) and how it is possible to detect asymmetry in the far-field. These goals can be achieved at once by incorporating into the setup an additional symmetric nanostructure which, under illumination by propagating fields, both produces a strong near-field coupling with the chiral film and scatters radiation into the far-field. To prove this, as shown in Fig.1(d), we equip the above setup with a transparent conductor nanowire of radius $a= 30 \: {\rm nm}$ lying on the chiral film and embedded into the substrate (the inset displays the nanowire permittivity $\epsilon_c$). We choose as incident propagating MSF, $F^{\rm (ei)}$, a mirror symmetric pair of plane waves whose electric field is 
\begin{equation}
{\bf{E}}^{\left( {\rm ei} \right)}  = E_0 \left[ {e^{i{\bf{K}} \cdot {\bf{r}}}  + e^{i \left(\mathcal{R} {\bf{K}} \right) \cdot {\bf{r}}} \mathcal{R}} \right] \left( { e_\parallel  {\bf{u}}_\parallel + e_ \bot  {\bf{u}}_ \bot     } \right),
\end{equation}
where $E_0$ is the amplitude of one of the plane waves, ${\bf{K}} = k_0 \sqrt {\varepsilon _1 } \left[ {\sin \Theta {\bf{e}}_x  + \cos \Theta \left( {\sin \Phi {\bf{e}}_y  + \cos \Phi {\bf{e}}_z } \right)} \right]$ is its wavevector along the direction of polar angles $(\Theta,\Phi)$, ${\bf{u}}_ \bot   = {\bf{K}} \times {\bf{e}}_x /\left| {{\bf{K}} \times {\bf{e}}_x } \right|$ and ${\bf{u}}_\parallel   = {\bf{u}}_ \bot   \times {\bf{K}}/\left| {{\bf{u}}_ \bot   \times {\bf{K}}} \right|$ are the transverse and parallel (to the nanowire) unit vectors orthogonal to $\bf K$ (see Fig.1(d)) and $e_ \bot  ,e_\parallel$ are dimensionless polarization coefficients satisfying $\left| {e_ \bot  } \right|^2  + \left| {e_\parallel  } \right|^2  = 1$. The field $F^{(\rm ei)}$ produces reflected and transmitted fields $F^{(\rm er)}$, $F^{(\rm et)}$ while the nanowire creates a radially outgoing field $F^{(\rm ci)}$ in the substrate in turn producing reflected and transmitted fields $F^{(\rm cr)}$, $F^{(\rm ct)}$ and they can be evaluated by the matching conditions at the nanowire surface (see Sec.S3 of Supplementary Material). Note that momentum conservation along the $x-$axis imposes the factor $e^{iK_x x}$ to the overall field whose relevant spatial dependence is only on $y,z$. The nanowire field $F^{(\rm ci)}$ is the superposition of TM and TE cylindrical modes with definite angular momentum along the $x-$axis, $e^{iK_x x} e^{im\varphi } H_m^{\left( 1 \right)} \left( {\sqrt {k_0^2 \varepsilon _1  - K_x^2 } \rho } \right)p^{\rm J} \left( m \right)$, where $\rm J=\{TM,TE\}$, the polar coordinates are coaxial with the nanowire (i.e. $y = \rho \cos \varphi$,  $z =  - a + \rho \sin \varphi$), $m$ is the topological charge proportional to the angular momentum, $H_m^{(1)}$ is the Hankel function of the first kind and $p^{\rm J}$ are modes expansion coefficients. The radiation scattered by the nanowire into the superstrate is decribed by the field $F^{\rm (ct)}$ whose far-field asymptotic expansion enables the evaluation of the differential cross length 
\begin{equation}
\Lambda \left( \psi  \right) = \frac{1}{{I^{\left( {\rm ei} \right)} }}\frac{{dP}}{{d\psi dx}}
\end{equation}
where, polar coordinates have been introduced (according to $y = r\cos \psi$, $z = r\sin \psi$, $0< \psi < \pi$), $dP$ is the power flowing through the asymptotical ($k_0 r \rightarrow \infty$) cylindrical surface $r d\psi dx$ and $I^{\left( {\rm ei} \right)}$ is the intensity of one incident plane wave (see Sec.S4 of Supplementary Material).

\begin{figure} \label{Fig3}
\center
\includegraphics[width=0.5\textwidth]{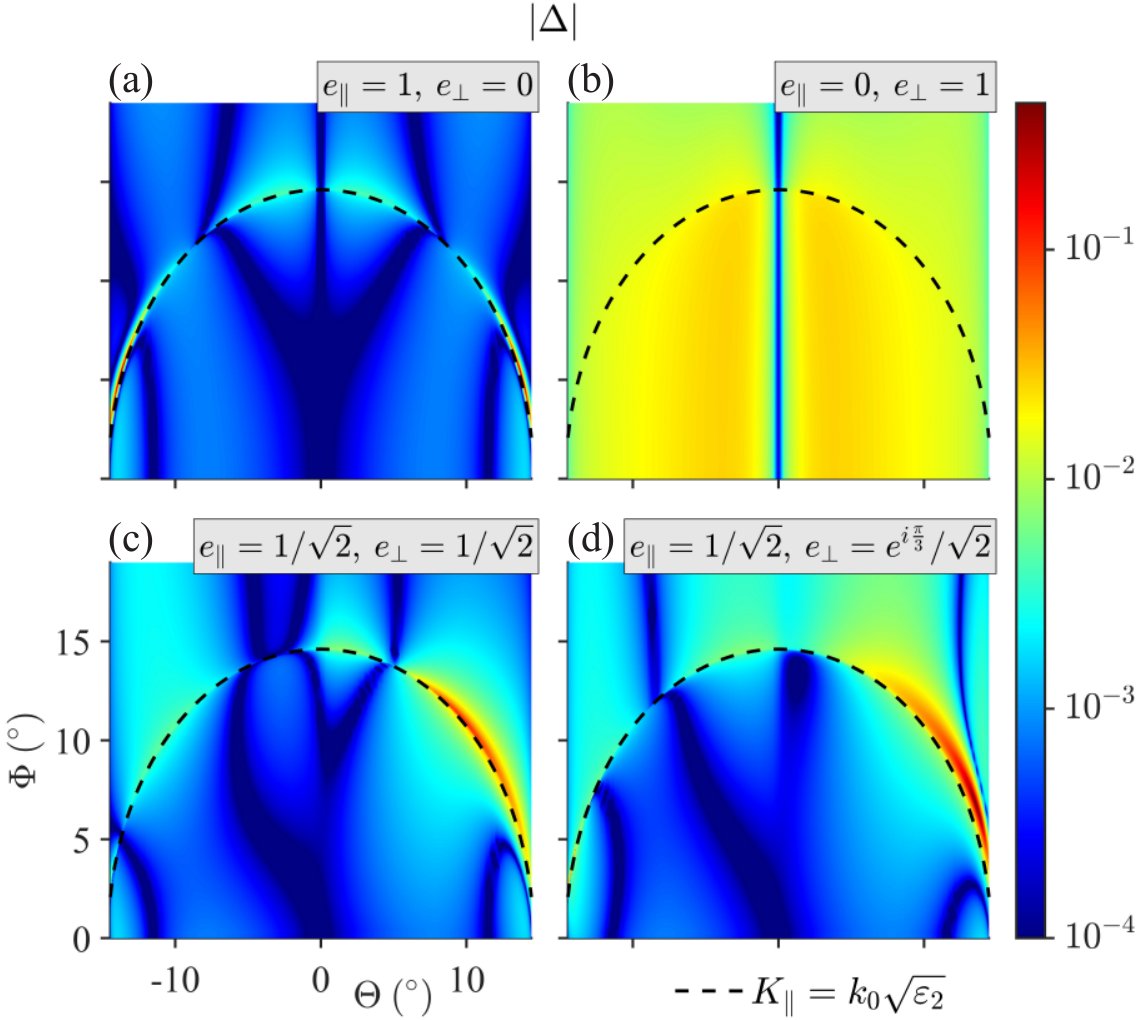}
\caption{Modulus of the scattered radiation dissymmetry factor $\Delta$ (see Eq.(\ref{rdf})) as function of the angles $\Theta,\Phi$ for various plane waves polarization states at molecular resonance $\lambda = 6.25 \: {\rm \mu m}$. The black dashed lines locate the threshold of total internal reflection (occurring above the curve) of the impinging plane waves.}
\end{figure}

Due to the deep subwavelength nanowire size ($a \ll \lambda$), large momentum photons ($k_\parallel > k_0$) dominate the field $F^{(\rm ci)}$ so that the reflected field $F^{(\rm cr)}$ has marked spatial asymmetry because of MOA (see the above discussion); in turn this implies that $F^{(\rm ci)}$ is asymmetric as well since the nanowire is also self-consistently stimulated by $F^{(\rm cr)}$ (in addition to $F^{(\rm ei)} + F^{(\rm er)}$) and consequently the modes with opposite angular momentum are not equally excited, i.e. $\left| {p^{\rm J} \left( m \right)} \right| \ne \left| {p^{\rm J} \left( { - m} \right)} \right|$. Each of these modes contributes to $F^{(\rm ct)}$ and to the radiation scattered into the superstrate (due to their low momentum photons $k_\parallel < k_0$) so that, by far-field interference, their unbalanced angular excitation produces an asymmetric scattered radiation pattern (see Fig.1(d)). These features are illustrated in Fig.2 for the specifc excitation parameters $\Theta = 13.48^\circ$, $\Phi = 11.39^\circ$, $e_\parallel = 1/\sqrt{2}$ and $e_\perp = e^{i \pi/3} /\sqrt{2}$ (see below for the directional and polarization analysis). In Fig.1(a) we plot the spatial dissymmetry factor 
\begin{equation} \label{sdf}
\delta \left| {\bf{E}} \right|\left( {\bf{r}} \right) = 2\frac{{\left| {{\bf{E}}\left( {\bf{r}} \right)} \right| - \left| {{\bf{E}}\left( {\mathcal{R}{\bf{r}}} \right)} \right|}}{{\left| {{\bf{E}}\left( {\bf{r}} \right)} \right| + \left| {{\bf{E}}\left( {\mathcal{R}{\bf{r}}} \right)} \right|}}
\end{equation}
of the overall substrate field ${\bf E} = {\bf E}^{\rm (ei) } + {\bf E}^{\rm (er) } + {\bf E}^{\rm (ci) } + {\bf E}^{\rm (cr) }$ at molecular resonance ($\lambda = 6.25 \: \mu {\rm m}$), cleary displaying the MOA induced asymmetry of the near-field surrounding the nanowire (black circle). In Fig.1(b) we plot the moduli of the nanowire modes coefficients ${p^{\rm TM} \left( m \right)}$ and ${p^{\rm TE} \left( m \right)}$ pertaining the situation of Fig.1(a) and the unequal excitation of the most relevant TE modes with opposite topological charges $m=-1$ and $m=+1$ is apparent; note that the nanowire excitation is accurately described by a small number of modes as a consequence of its small transverse size ($a \ll \lambda$). In Fig.1(c) we show the polar plot of the scattering differential cross length $\Lambda$  at molecular resonance (solid black line, corresponding to the situation of panels (a) and (b)) remarkably displaying the angular asymmetry of the radiation pattern. In order to characterize such angular asymmetry and to explicitly relate it to its source, the film chirality, in Fig.2(c) we also plot $\Lambda$ at two additional wavelengths (dotted red and dashed blue lines), clearly revealing that the asimmetry of the radiation pattern descreases by moving away from molecular resonace (see also panels  (b) and (c) of Fig.1), as expected; note that the strength of the radiation pattern is minimum at molecular resonance as a consequence of film absorption.

\begin{figure} \label{Fig4}
\center
\includegraphics[width=0.5\textwidth]{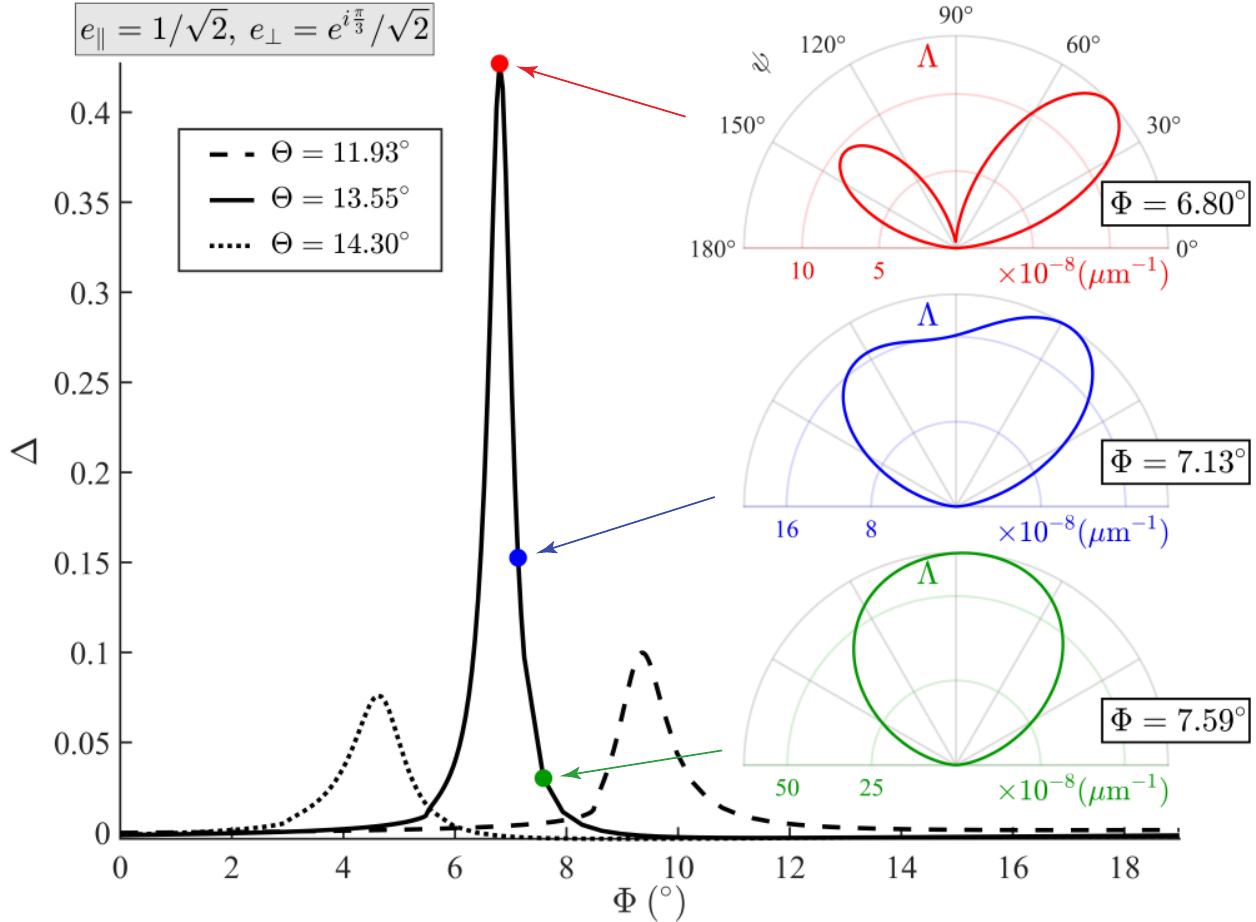}
\caption{Profiles of $|\Delta|$ extracted from Fig.3(d) versus $\Phi$ at three specific values of $\Theta$ around the lateral peak. The three insets display the scattering differential cross length $\Lambda$ corresponding to the three reported values of $\Phi$. 
}
\end{figure}

The nanowire-film geometry is such that plane waves direction and polarization have a marked impact on the radiation pattern asymmetry which can be quantitatively measured by the dissymmetry factor
\begin{equation} \label{rdf}
\Delta  = 2\frac{{\Lambda _ +   - \Lambda _ -  }}{{\Lambda _ +   + \Lambda _ -  }}
\end{equation}
where $\Lambda _ +   = \int\limits_0^{\pi /2} {d\psi \Lambda \left( \psi  \right)}$ and $\Lambda _ -   = \int\limits_{\pi /2}^\pi  {d\psi \Lambda \left( \psi  \right)}$ are the cross lengths of the radiation portions scattered into the lateral right and left sides of the superstrate, respectively. In Fig.3 we plot the modulus of $\Delta$ as function of the angles $\Theta,\Phi$ for various plane waves polarization states at molecular resonance $\lambda = 6.25 \: {\rm \mu m}$. We consider the angular range $\left| \Theta  \right| < \arcsin \sqrt {\varepsilon _2/\varepsilon _1}  \cong {\rm{14}}{\rm{.6^\circ }}$ since outside it the transmitted field $F^{\rm (ct)}$ comprises only evanescent photons owing to their definite momentum $K_x$ along the nanowire direction (due to the global factor $e^{iK_x x}$) and no radiation scattering occurs; we also consinder $\Phi >0$ for obvious symmetry reasons. In addition, it is worth locating the curve $K_\parallel /k_0  =  \sqrt {\varepsilon _2 } = 1$ (dashed black line) above which the incident plane waves undergo total internal reflection (since $\varepsilon_1 > \varepsilon_2$), as it is apparent from Fig.3 where the curve manifestly separates two different regimes in all the four cases. In fact along this curve the mixing coefficients $R_{\rm SP}$ and $R_{\rm PS}$ have a pronounced peak (see Fig.1(c)) so that MOA shows up also for the pure plane waves field and the  spatial asymmetry of $F^{(\rm er)}$ further enhances the overall asymmetric excitation of the nanowire. In Figs.3(a) and (b) we plot $|\Delta|$ for the parallel ($\parallel$) and transverse ($\bot$) plane waves polarizations, respectively, whose profiles are evidently even around $\Theta =0^\circ$ since $\Theta \rightarrow - \Theta$ merely amounts to exchanging the two plane waves. Note that the $\parallel$ polarization globally provides a smaller radiation dissymetry ($\simeq 10^{-3}$) than the $\bot$ one ($\simeq 10^{-2}$) and this is due to the fact that the overall cross length $\Lambda_+ + \Lambda_-$ is generally larger in the first case where the nanowire excitation is more effective since the dominant field component is along nanowire $x$- direction (see Fig.1(d)). In Fig.3(c) we consider the case of linear polarization with equal $\parallel$ and $\bot$ components and the enhancement of the scattered radiation asymmetry around the total internal reflection threshold curve (black dashed curve) is evident with a lateral peak around $\Theta = 10^\circ$ where $|\Delta| \simeq 10^{-1}$. Such peak corresponds to the smaller values of the the overall scattering cross length $\Lambda_+ + \Lambda_-$ resulting from the interference between the effects of the $\parallel$ and $\bot$ plane waves components. The same features characterize the case of the elliptical polarization state with $\pi/3$ phase difference between the $\parallel$ and $\bot$ components reported in Fig.3(d) where the peak is furher strengthened. 

In order to better appreciate such results, in Fig.4 we plot three sections of the dissymmetry factor $|\Delta|$ reported in Fig.3(d) corresponding to three different values of $\Theta$ around the peak. Note that the central profile corresponding to $\Theta = 13.55^\circ$ (solid balck curve) displays a peak maximum of about $0.43$ which is a giant asimmetry considering that it is entirely due to molecular chirality of a $50 \: {\rm nm}$ thick film. Such maximum asymmetry is dramatically confirmed by the profile of the corresponding differential cross length $\Lambda$ reported in the upper inset of Fig.4 (solid red line) from which the marked difference between the right and left sides of the scattered radiation pattern is evident. The other two insets of Fig.4 report the differential cross length $\Lambda$ (solid blue and green lines) pertaining two excitation states away from the peak maximum and they pictorially show the corresponding decrease of the radiation pattern asymmetry. Note that from the bottom to the upper inset of Fig.4 the overall strength of the radiation pattern decreases, in agreement with the above discussion about the interferometric mechanism that enhances $|\Delta|$ when the overall cross length is minimum.

In conclusion we have shown that marked spatial asymmetry can be detected in the radiation scattered by a nanowire coupled to a chiral molecular nanofilm when illuminated by mirror symmetric radiation. The occurrence of spatial asymmetry in the considered geometrically symmetric setup is physically due to electromagnetic parity indefiniteness characterizing chiral media whose chiroptical manifestation (mirror optical activity) 
is primarly a near-field effect. The nanowire plays two pivotal roles since it both generates the near-field necessary to trigger mirror optical activity and it literally carries spatial asymmetry from the near-field to the far-field. The peculiar behavior of plane waves undergoning total internal reflection provides, in the considered excitation scheme, further enhancemet of scattering asymmetry which can ultimately reach macroscopic levels, largely exceeding other kinds of chiroptical differential signals as those provided by circular dichrosim. Such giant asymmetric reponse and the simplicity of the setup, which is amenable to extensive generalizations, suggest that the here considered chiroptical effect could have a dramatic impact on conceiving ultra-efficient schemes for enantiomeric discrimantion of picogram
quantities of chiral molecules embedded in a nanofilm.

{\bf ACKNOWLEDGEMENTS} 

A.C acknowledges PRIN 2017 PELM (grant number 20177PSCKT).

\end{document}


\title{Asymmetric scattering of mirror symmetric radiation from nanostructures coupled to chiral films: Supplementary Material}
\author{A. Ciattoni$^1$}
\email{alessandro.ciattoni@spin.cnr.it}
\affiliation{$^1$CNR-SPIN, c/o Dip.to di Scienze Fisiche e Chimiche, Via Vetoio, 67100 Coppito (L'Aquila), Italy}
\date{\today}

\begin{abstract}

This document provides supporting information to ”Asymmetric scattering of mirror symmetric radiation from nanostructures coupled to chiral films”. We here review parity indefiniteness of the electromagnetic field in isotropic and homogeneous chiral media (Sec. S1) together with the analytical description of the interaction between radiation and a chiral slab, with emphasis on mirror optical activity (Sec. S2). Besides we provide analytical treatment of the scattering of a mirror symmetric pair of plane waves by a nanowire lying on a chiral slab (Sec. S3) and we introduce the differential cross length of the process (Sec. S4). 
\end{abstract}

\maketitle
\renewcommand\theequation{S\arabic{equation}}
\renewcommand\thefigure{S\arabic{figure}}
\renewcommand\thesection{S\arabic{section}}
\renewcommand\thesubsection{S\arabic{section}.\arabic{subsection}}
\count\footins = 1000

\section{Homogeneous Chiral Media and Electromagnetic Parity Indefiniteness}
We consider monochromatic electromagnetic fields ${\mathbfcal{E}}\left( {{\bf{r}},t} \right) = {\mathop{\rm Re}\nolimits} \left[ {{\bf{E}}\left( {\bf{r}} \right)e^{ - i\omega t} } \right]$, ${\mathbfcal{H}}\left( {{\bf{r}},t} \right) = {\mathop{\rm Re}\nolimits} \left[ {{\bf{H}}\left( {\bf{r}} \right)e^{ - i\omega t} } \right]$ and we adopt the notation for the complex amplitudes
\begin{equation} \label{F}
\def\arraystretch{1.3}
F = \left( {\begin{array}{@{\mkern0mu} c @{\mkern0mu}}
   {\bf{E}}  \\
   {Z_0 {\bf{H}}}  \\
\end{array}} \right)
\end{equation}
where $Z_0 = \sqrt{\mu_0/\epsilon_0}$ is the vacuum impedance. Maxwell equations in a homogenous and isotropic non-magnetic chiral medium are
\begin{equation} \label{Maxwell}
\def\arraystretch{1.3}
\left( {\begin{array}{@{\mkern0mu} c c @{\mkern0mu}}
   {\nabla  \times } & 0  \\
   0 & {\nabla  \times }  \\
\end{array}} \right)F = ik_0 \left( {\begin{array}{@{\mkern0mu} c c @{\mkern0mu}}
   {i\kappa } & 1  \\
   { - \varepsilon } & {i\kappa }  \\
\end{array}} \right)F
\end{equation}
where $k_0 = \omega /c$ is the vacuum wavenumber, $\varepsilon$ is the relative permittivity  and $\kappa$ is the Pasteur parameter. The parameter $\kappa$ controls the coupling between electric and magnetic responses of the medium and, therefore, the strength of its chirality. In the relevant case of a molecular medium where the chiral molecules are dispersed in a dielectric matrix medium, the permittivity and chirality parameter can be modelled as
\begin{eqnarray} \label{chi_disp}
 \varepsilon  &=& \varepsilon _{\rm b}  - \gamma_{\rm mol} \left( {\frac{1}{{\hbar \omega  - \hbar \omega _0  + i\Gamma }} - \frac{1}{{\hbar \omega  + \hbar \omega _0  + i\Gamma }}} \right), \nonumber \\ 
 \kappa  &=& \beta_{\rm mol} \left( {\frac{1}{{\hbar \omega  - \hbar \omega _0  + i\Gamma }} + \frac{1}{{\hbar \omega  + \hbar \omega _0  + i\Gamma }}} \right) 
\end{eqnarray}
where $\varepsilon _b$ is the background refractive index and the coefficients $\gamma_{\rm mol}$ and $\beta_{\rm mol}$ are the amplitudes of absorptive and chiral properties; $\omega_0 = 2\pi c / \lambda_0$ corresponds to the wavelength $\lambda_0$ of a molecular absorption resonance with broadening determined by the damping $\Gamma$. Relations (\ref{chi_disp}) can be derived from the quantum equation of motion for the density matrix of a chiral molecule \cite{Govor1} whose lack of definite parity enables the coupling between its electric and magnetic dipole moments in turn responsible for the magneto-electric coupling contained in the left hand side of Eqs.(\ref{Maxwell}).

In the present work we focus on chiroptical phenomena occurring in the infrared and we have chosen alanine as a specific molecular species for numerical calculations since it exhibits marked vibrational circular dichroism \cite{Tulioo,Jahnig}. Accordingly we hereafter set $\varepsilon_{\rm b} = 1.3$, $\rm \gamma_{mol} = 9.8 \cdot 10^{-4} \: eV$, $\rm \beta_{mol} = 4.68 \cdot 10^{-5} \: eV$, $\rm \hbar \omega_0 = 1.98 \cdot 10^{-1} \: eV$ (corresponding to the absorption resonance wavelength $\rm \lambda_0 = 6.25 \: \mu m$) and $\rm \Gamma = 1.6 \cdot 10^{-3} \: eV$ in order to emulate the resonant chiral response of an alanine enantiomer resulting from its vibrational mode at $\rm 1600 \: cm^{-1}$ (the two enantiomers have $\beta_{\rm mol}$, and hence $\kappa$, with opposite signs).

We now consider geometric mirror reflections and, without loss of generality, we focus on the reflection through the $xz$ plane, namely ${\bf{r}}' ={\mathcal{R}} {\bf{r}}$ where ${\mathcal{R}} = {{\bf{e}}_x {\bf{e}}_x - {\bf{e}}_y {\bf{e}}_y  + {\bf{e}}_z {\bf{e}}_z }$ and the dyadic notation $( {{\bf{ab}}} ){\bf{c}} = \left( {{\bf{b}} \cdot {\bf{c}}} \right) {\bf{a}}$ is hereafter adopted. Due to the respective polar and axial nature of the electric and magnetic fields, the mirror image of the field is  
\begin{equation}
\def\arraystretch{1.3}
F'\left( {\bf{r}} \right) = \left( {\begin{array}{@{\mkern0mu} c c @{\mkern0mu}}
   \mathcal{R} & 0  \\
   0 & { - \mathcal{R}}  \\
\end{array}} \right)F\left( {\mathcal{R}{\bf{r}}} \right)
\end{equation}
and, after setting 
\begin{equation}
F^{\rm S} = \frac{1}{2}\left( {F + F'} \right), \quad  
F^{\rm A} = \frac{1}{2}\left( {F - F'} \right)
\end{equation}
for its symmetric and antisymmetric parts, we are led to consider three kinds of fields:
\begin{eqnarray}
{\rm mirror \: symmetric \: field \: (MSF)}  &:& F^{\rm S}\ne 0, \: F^{\rm A}=0, \nonumber \\
{\rm mirror \: antisymmetric \: field \: (MAF)}  &:& F^{\rm S} = 0, \: F^{\rm A} \ne 0, \nonumber \\
{\rm indefinite \: parity \: field \: (IPF)}  &:& F^{\rm S} \ne 0, \: F^{\rm A} \ne 0.
\end{eqnarray}
In plain English, a MSF coincides with its mirror image ($F=F'$), a MAF is equal to the opposite of its mirror image ($F=-F'$) whereas an IPF has no geometric relation with its mirror image and it is asymmetric. As discussed in Ref.\cite{Ciatt1} the field $F$ existing in a homogeneous chiral medium (i.e. satisfying Eqs.(\ref{Maxwell})) is always an IPF since its mirror image $F'$  is easily found to satisfy the Maxwell equations
\begin{equation}
\def\arraystretch{1.3}
\left( {\begin{array}{@{\mkern0mu} c c @{\mkern0mu}}
   {\nabla  \times } & 0  \\
   0 & {\nabla  \times }  \\
\end{array}} \right)F' = ik_0 \left( {\begin{array}{@{\mkern0mu} c c @{\mkern0mu}}
   { - i\kappa } & 1  \\
   { - \varepsilon } & { - i\kappa }  \\
\end{array}} \right)F',
\end{equation}
i.e. $F'$ is an admissible field in the opposite enantiomeric medium, so that the possibilities $F=F'$ and $F=-F'$ are ruled out. We refer to this physical fact as parity indefiniteness of the electromagnetic field in homogeneous chiral media. Evidently, in a achiral medium where $\kappa =0$, MSFs, MAFs and IPFs are all separately allowed fields.

\begin{figure}
\begin{minipage}{.55\columnwidth}
\centering
\includegraphics[width = 0.95\linewidth]{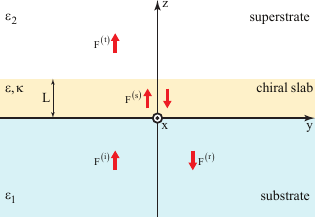}
\end{minipage}\hfill
\begin{minipage}{.45\columnwidth}
\begin{eqnarray}
   {{\bf{r}}_\parallel   = x{\bf{e}}_x  + y{\bf{e}}_x }       
   && \quad {\rm parallel \: position \: vector}        \nonumber \\
   {{\bf{k}}_\parallel   = k_x {\bf{e}}_x  + k_y {\bf{e}}_x } 
   && \quad {\rm parallel \: wavenumber}                \nonumber \\
   k_{1z}  = \sqrt {k_0^2 \varepsilon _1  - k_\parallel ^2 } 
   && \quad {\rm substrate \: longitudinal \: wavenumber}  \nonumber \\ 
   k_{2z}  = \sqrt {k_0^2 \varepsilon _2  - k_\parallel ^2 } 
   && \quad {\rm superstrate \: longitudinal \: wavenumber}  \nonumber \\
   {\bf{u}}_{\rm S}  = {\bf{e}}_z  \times \frac{{{\bf{k}}_\parallel  }}{{k_\parallel  }} 
   && \quad {\rm S \: parallel \: unit \: vector}  \nonumber \\
   {\bf{u}}_{\rm P}  = \frac{{{\bf{k}}_\parallel  }}{{k_\parallel  }}  
   && \quad {\rm P \: parallel \: unit \: vector}  \nonumber \\
   {n = \frac{{\sqrt {\varepsilon \kappa ^2 } }}{\kappa }} 
   && \quad {\rm chiral \: refractive \: index}  \nonumber 
\end{eqnarray}
\end{minipage}
\caption{{\bf Left:} Geometry of the chiral slab surrounded by a dielectric achiral environment made up of a substrate and a superstrate. Red arrows sketch the field configuration. {\bf Right:} List of quantities relevant for the present analysis}
\end{figure}

\section{Scattering by a chiral slab and Mirror Optical Activity}
With reference to the left side of Fig.S1, we here review the scattering of a monochromatic electromagnetic field by a chiral homogeneous and isotropic medium of permittivity $\varepsilon$ and chiral parameter $\kappa$ (see Sec.S1) filling the slab $0<z<L$, surrounded by a substrate in $z<0$ and a superstrate in $z>L$ of permittivities $\varepsilon_1$ and $\varepsilon_2$, respectively. For the sake of clarity we summarize in the right side of Fig.S1 the relevant quantities of our analysis. We here resort to the angular spectrum representation technique and we set 
\begin{equation}
\def\arraystretch{1.3}
F  = G_j^\sigma  \left[ {\left( {\begin{array}{@{\mkern0mu} c @{\mkern0mu}}
   {U_{\rm S} }  \\
   {U_{\rm P} }  \\
\end{array}} \right)} \right] \equiv \int {d^2 {\bf{k}}_\parallel  } e^{i{\bf{k}}_\parallel   \cdot {\bf{r}}_\parallel  } e^{i\sigma k_{jz} z} \left( {\begin{array}{@{\mkern0mu} c c @{\mkern0mu}}
 \displaystyle  {{\bf{u}}_{\rm S} } & \displaystyle  {\left( {{\bf{u}}_{\rm P}  - \frac{{k_\parallel  }}{{\sigma k_{jz} }}{\bf{e}}_z } \right)}  \\
 \displaystyle   { - \frac{{\sigma k_{jz} }}{{k_0 }}\left( {{\bf{u}}_{\rm P}  - \frac{{k_\parallel  }}{{\sigma k_{jz} }}{\bf{e}}_z } \right)} & \displaystyle  {\frac{{k_0 \varepsilon _j }}{{\sigma k_{jz} }}{\bf{u}}_{\rm S} }  \\
\end{array}} \right)\left( {\begin{array}{@{\mkern0mu} c @{\mkern0mu}}
   {U_{\rm S} }  \\
   {U_{\rm P} }  \\
\end{array}} \right)
\end{equation}
for the field $F$ (see Eq.(\ref{F})) propagating in the achiral medium $j=1,2$, with direction $\sigma = \pm 1$ along the $z$-axis, and with $\rm S$ and $\rm P$ amplitudes ${U_{\rm S} } \left( {{\bf{k}}_\parallel  } \right)$ and ${U_{\rm P} } \left( {{\bf{k}}_\parallel  } \right)$ in the  parallel-momentum space. In full generality, the incident (i) field in the substrate $z<0$ is 
\begin{equation} \label{i}
\def\arraystretch{1.3}
F^{\rm (i)} = G_1^ +  \left[ {\left( {\begin{array}{@{\mkern0mu} c @{\mkern0mu}}
   {U_{\rm S}^{\left(\rm i \right)} }  \\
   {U_{\rm P}^{\left(\rm i \right)} }  \\
\end{array}} \right)} \right]
\end{equation}
where ${U_{\rm S}^{\left(\rm i \right)} } \left( {{\bf{k}}_\parallel  } \right)$ and ${U_{\rm P}^{\left(\rm i \right)} } \left( {{\bf{k}}_\parallel  } \right)$ are the $\rm S$ and $\rm P$ amplitudes of the impinging photon of parallel momentum ${\bf k}_\parallel$. The reflected (r) and transmitted (r) fields, in the regions $z<0$ and $z>L$, respectively, are
\begin{equation} \label{rt}
\def\arraystretch{1.3}
F^{\rm (r)} = G_1^ -  \left[ {\left( {\begin{array}{@{\mkern0mu} c c @{\mkern0mu}}
   {R_{\rm SS} } & {nR_{\rm SP} }  \\
   {nR_{\rm PS} } & {R_{\rm PP} }  \\
\end{array}} \right)\left( {\begin{array}{@{\mkern0mu} c @{\mkern0mu}}
   {U_{\rm S}^{\left(\rm i \right)} }  \\
   {U_{\rm P}^{\left(\rm i \right)} }  \\
\end{array}} \right)} \right],\quad
F^{\rm (t)} = G_2^ +  \left[ {\left( {\begin{array}{@{\mkern0mu} c c @{\mkern0mu}}
   {T_{\rm SS} } & {nT_{\rm SP} }  \\
   {nT_{\rm PS} } & {T_{\rm PP} }  \\
\end{array}} \right)\left( {\begin{array}{@{\mkern0mu} c @{\mkern0mu}}
   {U_{\rm S}^{\left(\rm i \right)} }  \\
   {U_{\rm P}^{\left(\rm i \right)} }  \\
\end{array}} \right)} \right], 
\end{equation}
whose $\rm S$ and $\rm P$ amplitudes are obtained by the electromagnetic matching conditions at $z=0$ and $z=L$ with the slab field $F^{\rm (s)}$ (see left side of Fig.S1) \cite{Ciatt1}. The scattering by the slab is fully described by the coefficients $R_{\rm IJ} \left( {k_\parallel  ,\kappa ^2 } \right)$ and $T_{\rm IJ} \left( {k_\parallel  ,\kappa ^2 } \right)$ which are invariant under rotations around the $z-$axis and under chirality reversal $\kappa \rightarrow -\kappa$ (the sign of $\kappa$ is solely carried by the chiral refractive index $n$). Due to the circular polarization of the chiral slab eigenwaves, the S and P parts of the field are not independent (S/P coupling). Accordingly,  the S and P components of the reflected and transmitted fields depends on both $U_{\rm S}^{\left(\rm i \right)}$, $U_{\rm P}^{\left(\rm i \right)}$ due to the mixing  coefficients $R_{\rm SP}$, $R_{\rm PS}$, $T_{\rm SP}$, $T_{\rm PS}$ which are entirely due to chirality. In Fig.S2 we plot the $R_{\rm IJ}$ and $T_{\rm IJ}$ coefficients versus the normalized parallel wavenumber and the wavelength of a slab containing alanine (see Sec.S1), of thickness $L = 50 \: {\rm nm}$ and deposited on germanium ($\epsilon_1 = 15.731$) in vacuum ($\epsilon_2 = 1$). As opposed to the diagonal coefficients $R_{\rm II}$ and $T_{\rm II}$, the mixing coefficients $R_{\rm SP}$, $R_{\rm PS}$, $T_{\rm SP}$, $T_{\rm PS}$ are relevant only close to molecular resonance at $\lambda = 6.25 \: \mu {\rm m}$ in agreement with the chiral nature of the S/P coupling. It is worth noting that the coefficients $R_{\rm SP}$ and $T_{\rm SP}$ are not neglibible only along the evanescent part of the spectrum $k_\parallel > k_0$ where the subwvalegth spatial features of the field enhance the molecular magnetic dipole together with the role played by chirality. In addition the mixing reflection coefficients $R_{\rm SP}$ and $R_{\rm PS}$ display a local maximum at $k_\parallel = k_0$ (which is the threshold of total internal reflection of incident photons) since, at this parallel momentum, the photons transmitted in the superstrate travel along the slab surface ($k_{2z}=0$) and hence they have vanishing $\rm P$ component, this enhancing the effect of the chiral S/P coupling on the reflected field.

\begin{figure}
\includegraphics[width = \textwidth]{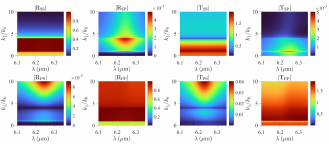}
\caption{Moduli of reflection and transmission coefficients of a chiral slab as functions of the normalized parallel wavevector $k_\parallel / k_0$ and of the wavelength $\lambda = 2 \pi c/\omega$.}
\end{figure}

The crucial point of the scattering we are considering is that the reflected and transmitted fields have always indefinite parity. To explicitly discuss this point we note that the symmetric and antisymmetric parts of the incident field of Eq.(\ref{i}) are
\begin{equation}
F^{\rm \left( i \right)S}  = G_1^ +  \left[ {U^{\left( \rm i \right)S} } \right], \quad 
F^{\rm \left( i \right)A}  = G_1^ +  \left[ {U^{\rm \left( i \right)A} } \right]
\end{equation}
whose parallel-momentum space components are
\begin{equation}
\def\arraystretch{1.3}
U^{\rm \left( i \right)S} \left( {{\bf{k}}_\parallel  } \right) = \frac{1}{2}\left( {\begin{array}{@{\mkern0mu} c @{\mkern0mu}}
   {U_{\rm S}^{\rm \left( i \right)} \left( {{\bf{k}}_\parallel  } \right) - U_{\rm S}^{\rm \left( i \right)} \left( {\mathcal{R}{\bf{k}}_\parallel  } \right)}  \\
   {U_{\rm P}^{\rm \left( i \right)} \left( {{\bf{k}}_\parallel  } \right) + U_{\rm P}^{\rm \left( i \right)} \left( {\mathcal{R}{\bf{k}}_\parallel  } \right)}  \\
\end{array}} \right),\quad 
U^{\rm \left( i \right)A} \left( {{\bf{k}}_\parallel  } \right) = \frac{1}{2}\left( {\begin{array}{@{\mkern0mu} c @{\mkern0mu}}
   {U_{\rm S}^{\rm \left( i \right)} \left( {{\bf{k}}_\parallel  } \right) + U_{\rm S}^{\rm \left( i \right)} \left( {\mathcal{R}{\bf{k}}_\parallel  } \right)}  \\
   {U_{\rm P}^{\rm \left( i \right)} \left( {{\bf{k}}_\parallel  } \right) - U_{\rm P}^{\rm \left( i \right)} \left( {\mathcal{R}{\bf{k}}_\parallel  } \right)}  \\
\end{array}} \right).
\end{equation}
where use has been made of the relations 
\begin{eqnarray}
{\bf{u}}_{\rm S} \left( {\mathcal{R}{\bf{k}}_\parallel  } \right) &=&  - \mathcal{R}{\bf{u}}_{\rm S} \left( {{\bf{k}}_\parallel  } \right), \nonumber \\ 
{\bf{u}}_{\rm P} \left( {\mathcal{R}{\bf{k}}_\parallel  } \right) &=& \mathcal{R}{\bf{u}}_{\rm P} \left( {{\bf{k}}_\parallel  } \right).
\end{eqnarray}
Evidently, $U^{\rm \left( i \right)S}$ and $U^{\rm \left( i \right)A}$ are independent amplitudes and the incident field can both have definite (MSF, MAF) or indefinite parity (IPF). On the other hand, the symmetric and antisymmetric parts of the reflected field of Eqs.(\ref{rt}) are
\begin{equation}
\def\arraystretch{1.3}
F^{\rm \left( r \right)S}  = G_1^ -  \left[ {\left( {\begin{array}{@{\mkern0mu} c c @{\mkern0mu}}
   R_{\rm SS} & 0  \\
   0 & \rm R_{PP}   \\
\end{array}} \right)U^{\rm \left( i \right)S}  + n\left( {\begin{array}{@{\mkern0mu} c c @{\mkern0mu}}
   0 & {R_{\rm SP} }  \\
   {R_{\rm PS} } & 0  \\
\end{array}} \right)U^{\rm \left( i \right)A} } \right], \quad F^{\rm \left( r \right)A}  = G_1^ -  \left[ {\left( {\begin{array}{@{\mkern0mu} c c @{\mkern0mu}}
   {R_{\rm SS} } & 0  \\
   0 & {R_{\rm PP} }  \\
\end{array}} \right)U^{\rm \left( i \right)A}  + n\left( {\begin{array}{@{\mkern0mu} c c @{\mkern0mu}}
   0 & {R_{\rm SP} }  \\
   {R_{\rm PS} } & 0  \\
\end{array}} \right)U^{\rm \left( i \right)S} } \right],
\end{equation}
vividly showing that $F^{\rm (r)}$ is always and IPF, no matter what is the parity of the incident field, since $F^{\rm \left( r \right)S}$ and $F^{\rm \left( r \right)A}$ can not independently vanish. Note that, in the limiting achiral case $R_{\rm SP}=R_{\rm PS}=0$, the reflected field has the same parity of the incident field, thus confirming that the parity braking of the scattered field is genuinely a chiro-optical effect produced by the slab chirality. In the special case of an incident MSF we have
\begin{eqnarray}
\def\arraystretch{1.3}
U^{\rm \left( i \right)S}  = \left( {\begin{array}{@{\mkern0mu} c @{\mkern0mu}}
   {U_{\rm S}^{\left(\rm i \right)} }  \\
   {U_{\rm P}^{\left(\rm i \right)} }  \\
\end{array}} \right)&,& \quad 
U^{\rm \left( i \right)A}  = 0, \nonumber \\
F^{\rm \left( r \right)S}  = G_1^ -  \left[ {\left( {\begin{array}{@{\mkern0mu} c c @{\mkern0mu}}
   {R_{\rm SS} } & 0  \\
   0 & {R_{\rm PP} }  \\
\end{array}} \right)\left( {\begin{array}{@{\mkern0mu} c @{\mkern0mu}}
   {U_{\rm S}^{\rm \left( i \right)} }  \\
   {U_{\rm P}^{\rm \left( i \right)} }  \\
\end{array}} \right)} \right]&,& \quad 
F^{\rm \left( r \right)A}  = G_1^ -  \left[ {n\left( {\begin{array}{@{\mkern0mu} c c @{\mkern0mu}}
   0 & {R_{\rm SP} }  \\
   {R_{\rm PS} } & 0  \\
\end{array}} \right)\left( {\begin{array}{@{\mkern0mu} c @{\mkern0mu}}
   {U_{\rm S}^{\rm \left( i \right)} }  \\
   {U_{\rm P}^{\rm \left( i \right)} }  \\
\end{array}} \right)} \right]
\end{eqnarray}
clearly displaying the asymmetry of the reflected field. The chiroptical effect where an incident symmetric field produces reflected and transmitted asymmetric fields is a consequence of the electromagnetic parity indefinitenss in chiral media (see Sec. S1) and it is referred to as mirror optical activity (MOA) \cite{Ciatt1}.

\section{Scattering of a mirror symmetric pair of plane waves by a nanowire in the substrate }
With reference to Fig.S3, we now consider the same setup as in Fig.S2 with an additional nanowire (cylinder) embedded in the substrate, lying on the chiral slab, of radius $a$ and permittivity $\varepsilon_{\rm c}$. In this work we consider the scattering of a pair of mirror symmetric planes waves from the cylinder/chiral slab and accordingly we set for the external (e) incident field in $z<0$
\begin{equation}
\def\arraystretch{1.3}
F^{\left( {\rm ei} \right)}  = E_0 \left[ {e^{i{\bf{K}} \cdot {\bf{r}}} \left( {\begin{array}{@{\mkern0mu} cc @{\mkern0mu}}
   1 & 0  \\
   0 & 1  \\
\end{array}} \right) + e^{i{\bf{K}} \cdot \mathcal{R}{\bf{r}}} \left( {\begin{array}{@{\mkern0mu} cc @{\mkern0mu}}
   \mathcal{R} & 0  \\
   0 & { - \mathcal{R}}  \\
\end{array}} \right)} \right]\left( {\begin{array}{@{\mkern0mu} c @{\mkern0mu}}
   {\left( {e_ \bot  {\bf{u}}_ \bot   + e_\parallel  {\bf{u}}_\parallel  } \right)}  \\
   {\sqrt {\varepsilon _1 } \left( {e_\parallel  {\bf{u}}_ \bot   - e_ \bot  {\bf{u}}_\parallel  } \right)}  \\
\end{array}} \right)
\end{equation}
where $E_0$ is the amplitude of one of the plane waves, ${\bf{K}} = k_0 \sqrt {\varepsilon _1 } \left[ {\sin \Theta {\bf{e}}_x  + \cos \Theta \left( {\sin \Phi {\bf{e}}_y  + \cos \Phi {\bf{e}}_z } \right)} \right]$ is its wavevector, ${\bf{u}}_ \bot   = {\bf{K}} \times {\bf{e}}_x /\left| {{\bf{K}} \times {\bf{e}}_x } \right|$, ${\bf{u}}_\parallel   = {\bf{u}}_ \bot   \times {\bf{K}}/\left| {{\bf{u}}_ \bot   \times {\bf{K}}} \right|$ are its transverse and parallel (to the cylinder) polarization unit vectors (see Fig.1 of the main text) and $e_\parallel$, $e_\bot$ are dimensionless polarization coefficients satysfing the relation $\left| {e_ \bot  } \right|^2  + \left| {e_\parallel  } \right|^2  = 1$. After some algebra, the external incident field can be casted as (see Eq.(\ref{i}))
\begin{equation}
\def\arraystretch{1.3}
F^{\left( {\rm ei} \right)}  = G_1^ +  \left[ {\left( {\begin{array}{@{\mkern0mu} c @{\mkern0mu}}
   {U_{\rm S}^{\left( {\rm ei} \right)} }  \\
   {U_{\rm P}^{\left( {\rm ei} \right)} }  \\
\end{array}} \right)} \right],
\end{equation}
where 
\begin{equation}
\def\arraystretch{1.3}
\left( {\begin{array}{@{\mkern0mu} c @{\mkern0mu}}
   {U_{\rm S}^{\left( {\rm ei} \right)} }  \\
   {U_{\rm P}^{\left( {\rm ei} \right)} }  \\
\end{array}} \right) = E_0 \frac{{\delta \left( {k_x  - K_x } \right)}}{{k_\parallel  }}\frac{{k_{1z} }}{{\sqrt {k_0^2 \varepsilon _1  - k_x^2 } }}\left(  {\begin{array}  
{@{\mkern0mu} cc @{\mkern0mu}}
  { - \frac{{\sqrt {\varepsilon _1 } k_y }}{{k_{1z} }}} & {\frac{{k_x }}{{k_0 }}}  \\
   {\frac{{k_x k_{1z} }}{{k_0^2 \sqrt {\varepsilon _1 } }}} & {\frac{{k_y }}{{k_0 }}}  \\
\end{array}} \right)\left[ {k_0 \delta \left( {k_y  - K_y } \right)\left( {\begin{array}{@{\mkern0mu} c @{\mkern0mu}}
   {e_ \bot  }  \\
   {e_\parallel  }  \\
\end{array}} \right) + k_0 \delta \left( {k_y  + K_y } \right)\left( {\begin{array}{@{\mkern0mu} c @{\mkern0mu}}
   {e_ \bot  }  \\
   { - e_\parallel  }  \\
\end{array}} \right)} \right],
\end{equation}
so that, the reflected and transmitted field in the absence of the cylinder are (see Eqs.(\ref{rt}) and Fig.S3)
\begin{equation}
\def\arraystretch{1.3}
F^{\left( {\rm er} \right)}  = G_1^ -  \left[ {\left( {\begin{array}{@{\mkern0mu} cc @{\mkern0mu}}
   {R_{\rm SS} } & {nR_{\rm SP} }  \\
   {nR_{\rm PS} } & {R_{\rm PP} }  \\
\end{array}} \right)\left( {\begin{array}{@{\mkern0mu} c @{\mkern0mu}}
   {U_{\rm S}^{\left( {\rm ei} \right)} }  \\
   {U_{\rm P}^{\left( {\rm ei} \right)} }  \\
\end{array}} \right)} \right],\quad F^{\left( {\rm et} \right)}  = G_2^ +  \left[ {\left( {\begin{array}{@{\mkern0mu} cc @{\mkern0mu}}
   {T_{\rm SS} } & {nT_{\rm SP} }  \\
   {nT_{\rm PS} } & {T_{\rm PP} }  \\
\end{array}} \right)\left( {\begin{array}{@{\mkern0mu} c @{\mkern0mu}}
   {U_{\rm S}^{\left( {\rm ei} \right)} }  \\
   {U_{\rm P}^{\left( {\rm ei} \right)} }  \\
\end{array}} \right)} \right].
\end{equation}
\begin{figure}
\includegraphics[width = 0.5\textwidth]{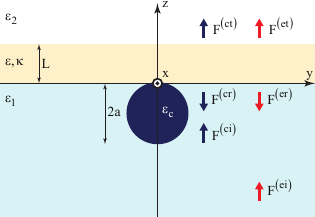}
\caption{Same setup as in Fig.S1 with the additional nanowire (cylinder) of radius $a$ and lying on the chiral slab in the substrate. The external (e) and cylinder (c) fields both have incident (i), reflected (r) and transmitted (t) parts. }
\end{figure}
Since the external field $F^{(\rm ei)}$ imposes the $e^{iK_x x}$ spatial factor to the whole electromagnetic field, the nanowire (c) field can be expanded in cylindrical harmonics in $z<0$ as \cite{Chew} 
\begin{equation} \label{Fci}
\def\arraystretch{1.3}
F^{\left( {\rm ci} \right)}  = \left\{ \begin{array}{ll}
 \frac{1}{{k_0^2 }}\left( {\begin{array}{@{\mkern0mu} cc @{\mkern0mu}}
   {k_0^2 \varepsilon _1  + \nabla \nabla  \cdot } & {ik_0 \nabla  \times }  \\
   { - ik_0 \varepsilon _1 \nabla  \times } & {k_0^2 \varepsilon _1  + \nabla \nabla  \cdot }  \\
\end{array}} \right)\left[ {{\bf{e}}_x e^{iK_x x} \sum\limits_{m =  - \infty }^{ + \infty } {e^{im\varphi } } H_m^{\left( 1 \right)} \left( {K_1 \rho } \right)\left( {\begin{array}{@{\mkern0mu} c @{\mkern0mu}}
   {p^{\rm TM} \left( m \right)}  \\
   {p^{\rm TE} \left( m \right)}  \\
\end{array}} \right)} \right], & \quad \quad \quad \rho  > a, \\ 
 &  \\
 \frac{1}{{k_0^2 }}\left( {\begin{array}{@{\mkern0mu} cc @{\mkern0mu}}
   {k_0^2 \varepsilon _c  + \nabla \nabla  \cdot } & {ik_0 \nabla  \times }  \\
   { - ik_0 \varepsilon _c \nabla  \times } & {k_0^2 \varepsilon _c  + \nabla \nabla  \cdot }  \\
\end{array}} \right)\left[ {{\bf{e}}_x e^{iK_x x} \sum\limits_{m =  - \infty }^{ + \infty } {e^{im\varphi } } J_m \left( {K_c \rho } \right)\left( {\begin{array}{@{\mkern0mu} c @{\mkern0mu}}
   {q^{\rm TM} \left( m \right)}  \\
   {q^{\rm TE} \left( m \right)}  \\
\end{array}} \right)} \right], & \quad \quad \quad \rho  < a, \\ 
 \end{array} \right. 
\end{equation}
where cylindrical coordinates $(\rho,\varphi)$ coaxial with the cylinder have been introduced according to $y = \rho \cos \varphi$, $z =  - a + \rho \sin \varphi$, $K_1  = \sqrt {k_0^2 \varepsilon _1  - K_x^2 }$ and $K_c  = \sqrt {k_0^2 \varepsilon _c  - K_x^2 }$ are the radial wavenumbers, $H_m^{(1)}$ and $J_m$ are the Hankel and Bessel functions of the first kind, $\left( {p^{\rm TM} \left( m \right),p^{\rm TE} \left( m \right)} \right)$ and $\left( {q^{\rm TM} \left( m \right),q^{\rm TE} \left( m \right)} \right)$ are the external and internal weights of the $m$-th cylindrical harmonics. By using the integral expression for the cylindrical harmonics \cite{Schwinger}
\begin{equation}
e^{im\varphi } H_m^{\left( 1 \right)} \left( {K_1 \rho } \right) = \left[ { - \frac{1}{{K_1 }}\left( {\frac{\partial }{{\partial y}} + i\frac{\partial }{{\partial z}}} \right)} \right]^m \frac{1}{\pi }\int {dk_y } e^{ik_y y} \frac{{e^{i\sqrt {K_1^2  - k_y^2 } \left| {z + a} \right|} }}{{\sqrt {K_1^2  - k_y^2 } }},
\end{equation}
the field $F^{\left( {\rm ci} \right)}$  can be casted, in $z<0$ and $\rho>a$, as (see Eq.(\ref{i}) and Fig.S3)
\begin{equation}
\def\arraystretch{1.3}
F^{\left( {\rm ci} \right)}  = G_1^ +  \left[ {\left( {\begin{array}{@{\mkern0mu} c @{\mkern0mu}}
   {U_{\rm S}^{\left( {\rm ci} \right)} }  \\
   {U_{\rm P}^{\left( {\rm ci} \right)} }  \\
\end{array}} \right)} \right],
\end{equation}
where 
\begin{equation} \label{Uci}
\def\arraystretch{1.3}
\left( {\begin{array}{@{\mkern0mu} c @{\mkern0mu}}
   {U_{\rm S}^{\left( {\rm ci} \right)} }  \\
   {U_{\rm P}^{\left( {\rm ci} \right)} }  \\
\end{array}} \right) = \frac{{\delta \left( {k_x  - K_x } \right)}}{{\pi k_\parallel  }}e^{ik_{1z} a} \left( {\begin{array}{@{\mkern0mu} cc @{\mkern0mu}}
   { -  \frac{\varepsilon _1 {k_y }}{{k_{1z} }}} & { - \frac{{k_x }}{{k_0 }}}  \\
   {\frac{{k_x k_{1z} }}{{k_0^2 }}} & { - \frac{{k_y }}{{k_0 }}}  \\
\end{array}} \right)\sum\limits_{m =  - \infty }^\infty  {\left( {  \frac{{-ik_y  + k_{1z} }}{{K_1 }}} \right)^m \left( {\begin{array}{@{\mkern0mu} c @{\mkern0mu}}
   {p^{\rm TM} \left( m \right)}  \\
   {p^{\rm TE} \left( m \right)}  \\
\end{array}} \right)} 
\end{equation}
so that, the reflected and transmitted field are (see Eqs.(\ref{rt}))
\begin{equation} \label{FcrFct}
\def\arraystretch{1.3}
F^{\left( {\rm cr} \right)}  = G_1^ -  \left[ {\left( {\begin{array}{@{\mkern0mu} cc @{\mkern0mu}}
   {R_{\rm SS} } & {nR_{\rm SP} }  \\
   {nR_{\rm PS} } & {R_{\rm PP} }  \\
\end{array}} \right)\left( {\begin{array}{@{\mkern0mu} c @{\mkern0mu}}
   {U_{\rm S}^{\left( {\rm ci} \right)} }  \\
   {U_{\rm P}^{\left( {\rm ci} \right)} }  \\
\end{array}} \right)} \right],\quad F^{\left( {\rm ct} \right)}  = G_2^ +  \left[ {\left( {\begin{array}{@{\mkern0mu} cc @{\mkern0mu}}
   {T_{\rm SS} } & {nT_{\rm SP} }  \\
   {nT_{\rm PS} } & {T_{\rm PP} }  \\
\end{array}} \right)\left( {\begin{array}{@{\mkern0mu} c @{\mkern0mu}}
   {U_{\rm S}^{\left( {\rm ci} \right)} }  \\
   {U_{\rm P}^{\left( {\rm ci} \right)} }  \\
\end{array}} \right)} \right].
\end{equation}
The above (e) and (c) fields satisfy Maxwell equations in their own regions and electromagnetic matching conditions at $z=0$ and $z=L$ (see Sec.S2). In order to set the electromagnetic matching conditions at the cylinder surface $\rho = a$ we note that the cylinder is excited both by the external fields $F^{\rm (ei)}$, $F^{\rm (er)}$ and by the cylinder reflected field $F^{\rm (cr)}$ and these fields admit the cylindrical harmonics expansions in $z<0$ and $\rho >a$ given by
\begin{eqnarray}
\def\arraystretch{1.3}
 F^{\left( {\rm ei} \right)}  + F^{\left( {\rm er} \right)}  &=& \frac{1}{{k_0^2 }}\left( {\begin{array}{@{\mkern0mu} cc @{\mkern0mu}}
   {k_0^2 \varepsilon _1  + \nabla \nabla  \cdot } & {ik_0 \nabla  \times }  \\
   { - ik_0 \varepsilon _1 \nabla  \times } & {k_0^2 \varepsilon _1  + \nabla \nabla  \cdot }  \\
\end{array}} \right)\left[ {{\bf{e}}_x e^{iK_x x} \sum\limits_{m =  - \infty }^{ + \infty } {e^{im\varphi } } J_m \left( {K_1 \rho } \right)\left( {\begin{array}{@{\mkern0mu} c @{\mkern0mu}}
   {Q^{\rm TM} \left( m \right)}  \\
   {Q^{\rm  TE} \left( m \right)}  \\
\end{array}} \right)} \right], \nonumber \\ 
 F^{\left( {\rm cr} \right)}  &=& \frac{1}{{k_0^2 }}\left( {\begin{array}{@{\mkern0mu} cc @{\mkern0mu}}
   {k_0^2 \varepsilon _1  + \nabla \nabla  \cdot } & {ik_0 \nabla  \times }  \\
   { - ik_0 \varepsilon _1 \nabla  \times } & {k_0^2 \varepsilon _1  + \nabla \nabla  \cdot }  \\
\end{array}} \right)\left\{ {{\bf{e}}_x e^{iK_x x} \sum\limits_{m =  - \infty }^{ + \infty } {e^{im\varphi } } J_m \left( {K_1 \rho } \right)\left[ {\sum\limits_{n =  - \infty }^\infty  {\left( { - 1} \right)^n U\left( {m + n} \right)\left( {\begin{array}{@{\mkern0mu} c @{\mkern0mu}}
   {p^{\rm TM} \left( n \right)}  \\
   {p^{\rm TE} \left( n \right)}  \\
\end{array}} \right)} } \right]} \right\}, \nonumber \\ 
\end{eqnarray}
where 
\begin{eqnarray}
\def\arraystretch{1.3}
 \left( {\begin{array}{@{\mkern0mu} c @{\mkern0mu}}
   {Q^{\rm TM} \left( m \right)}  \\
   {Q^{\rm TE} \left( m \right)}  \\
\end{array}} \right) &=& E_0 \frac{{k_0 }}{{K_1 }}e^{ - iK_z a} \left[ {\left( {\frac{{iK_y  + K_z }}{{K_1 }}} \right)^m \left( {\begin{array}{@{\mkern0mu} cc @{\mkern0mu}}
   1 & 0  \\
   0 & { - 1}  \\
\end{array}} \right) + \left( {\frac{{ - iK_y  + K_z }}{{K_1 }}} \right)^m \left( {\begin{array}{@{\mkern0mu} cc @{\mkern0mu}}
   1 & 0  \\
   0 & 1  \\
\end{array}} \right)} \right]\left( {\begin{array}{@{\mkern0mu} c @{\mkern0mu}}
   {\frac{1}{{\sqrt {\varepsilon _1 } }}e_\parallel  }  \\
   {e_ \bot  }  \\
\end{array}} \right) + \nonumber \\ 
 &+& E_0 \left( {\frac{{k_0 }}{{K_1 }}} \right)^3 \frac{{k_0 K_z }}{{K_x^2  + K_y^2 }}e^{iK_z a} \left[ {\left( {\frac{{iK_y  - K_z }}{{K_1 }}} \right)^m \left( {\begin{array}{@{\mkern0mu} cc @{\mkern0mu}}
   { - \frac{{K_y }}{{k_0 }}} & {\frac{{K_x }}{{k_0 }}}  \\
   {\frac{{K_x K_z }}{{k_0^2 }}} & { \frac{\varepsilon _1{K_y }}{{K_z }}}  \\
\end{array}} \right)\left( {\begin{array}{@{\mkern0mu} cc @{\mkern0mu}}
   {R_{\rm SS} } & {nR_{\rm SP} }  \\
   {nR_{\rm PS} } & {R_{\rm PP} }  \\
\end{array}} \right)_{k_\parallel   = \sqrt {K_x^2  + K_y^2 } } \left( {\begin{array}{@{\mkern0mu} cc @{\mkern0mu}}
   { - \frac{{\sqrt {\varepsilon _1 } K_y }}{{K_z }}} & {\frac{{K_x }}{{k_0 }}}  \\
   {\frac{{K_x K_z }}{{k_0^2 \sqrt {\varepsilon _1 } }}} & {\frac{{K_y }}{{k_0 }}}  \\
\end{array}} \right) + } \right. \nonumber \\ 
 &&\left. { + \left( {\frac{{ - iK_y  - K_z }}{{K_1 }}} \right)^m \left( {\begin{array}{@{\mkern0mu} cc @{\mkern0mu}}
   { - \frac{{K_y }}{{k_0 }}} & {\frac{{K_x }}{{k_0 }}}  \\
   { - \frac{{K_x K_z }}{{k_0^2 }}} & { - \varepsilon _1 \frac{{K_y }}{{K_z }}}  \\
\end{array}} \right)\left( {\begin{array}{@{\mkern0mu} cc @{\mkern0mu}}
   {R_{\rm SS} } & { - nR_{\rm SP} }  \\
   { - nR_{\rm PS} } & {\rm R_{PP} }  \\
\end{array}} \right)_{k_\parallel   = \sqrt {K_x^2  + K_y^2 } } \left( {\begin{array}{@{\mkern0mu} cc @{\mkern0mu}}
   { - \frac{{\sqrt {\varepsilon _1 } K_y }}{{K_z }}} & {\frac{{K_x }}{{k_0 }}}  \\
   {\frac{{K_x K_z }}{{k_0^2 \sqrt {\varepsilon _1 } }}} & {\frac{{K_y }}{{k_0 }}}  \\
\end{array}} \right)} \right]\left( {\begin{array}{@{\mkern0mu} c @{\mkern0mu}}
   {e_\parallel  }  \\
   {e_ \bot  }  \\
\end{array}} \right) \nonumber \\ 
 U\left( s \right) &=& \int {d^2 {\bf{k}}_\parallel  } \frac{{\delta \left( {k_x  - K_x } \right)}}{{\pi k_\parallel  }}e^{i2k_{1z} a} \frac{{k_0^3 }}{{k_{1z} K_1^2 }}\left( {\begin{array}{@{\mkern0mu} cc @{\mkern0mu}}
   { - \frac{{k_y }}{{k_0 }}} & {\frac{{k_x }}{{k_0 }}}  \\
   {\frac{{k_x k_{1z} }}{{k_0^2 }}} & { \frac{\varepsilon _1 {k_y }}{{k_{1z} }}}  \\
\end{array}} \right)\left( {\begin{array}{@{\mkern0mu} cc @{\mkern0mu}}
   {R_{\rm SS} } & {nR_{\rm SP} }  \\
   {nR_{\rm PS} } & {R_{\rm PP} }  \\
\end{array}} \right)\left( {\begin{array}{@{\mkern0mu} cc @{\mkern0mu}}
   { - \frac{{\varepsilon _1 k_y }}{{k_\parallel  }}} & { - \frac{{k_x k_{1z} }}{{k_0 k_\parallel  }}}  \\
   {\frac{{k_x k_{1z}^2 }}{{k_0^2 k_\parallel  }}} & { - \frac{{k_y k_{1z} }}{{k_0 k_\parallel  }}}  \\
\end{array}} \right)\left( {\frac{{ik_y  - k_{1z} }}{{K_1 }}} \right)^s. \nonumber \\ 
\end{eqnarray}
The coefficients $Q^{\rm TM},Q^{\rm TE}$ and the matrices $U(s)$ have been obtained, after some tediuos but straighforward algebra, by using the relation for the Bessel functions \cite{Schwinger}
\begin{equation}
\left[ {\frac{1}{{K_1 }}\left( {\frac{\partial }{{\partial y}} - i\frac{\partial }{{\partial z}}} \right)} \right]^m e^{im\varphi } J_m \left( {K_1 \rho } \right) = J_0 \left( {K_1 \rho } \right)
\end{equation}
and the fact that $J_n(0) = \delta_{n,0}$. Now the fields just outside and inside the cylinder surface are $\left[ F^{\left( {\rm ei} \right)}  + F^{\left( {\rm er} \right)} + F^{\left( {\rm ci} \right)}  + F^{\left( {\rm cr} \right)}\right]_{\rho = a^+}$ and $\left[ F^{\left( {\rm ci} \right)}  \right]_{\rho = a^-}$, respectively, so that, the continuity of the tangential electric and magnetic field components readily yield \cite{Chew}
\begin{equation} \label{system}
\def\arraystretch{1.3}
\left( {\begin{array}{@{\mkern0mu} c @{\mkern0mu}}
   {p^{\rm TM} \left( m \right)}  \\
   {p^{\rm TE} \left( m \right)}  \\
\end{array}} \right) = M\left( m \right)\left[ {\left( {\begin{array}{@{\mkern0mu} c @{\mkern0mu}}
   {Q^{\rm TM} \left( m \right)}  \\
   {Q^{\rm TE} \left( m \right)}  \\
\end{array}} \right) + \sum\limits_{n =  - \infty }^\infty  {\left( { - 1} \right)^n U\left( {m + n} \right)\left( {\begin{array}{@{\mkern0mu} c @{\mkern0mu}}
   {p^{\rm TM} \left( n \right)}  \\
   {p^{\rm TE} \left( n \right)}  \\
\end{array}} \right)} } \right]
\end{equation}
where the matching matrix is
\begin{eqnarray}
\def\arraystretch{1.3}
 M\left( m \right) &=&  - \left( {\begin{array}{@{\mkern0mu} cc @{\mkern0mu}}
   {K_x m\frac{{H_m^{\left( 1 \right)} \left( {K_1 a} \right)}}{{k_0 a}}\left( {\frac{{K_1^2 }}{{K_c^2 }} - 1} \right)} & {iK_1 H_m^{\left( 1 \right)} \left( {K_1 a} \right)\left[ { - \frac{{H_m^{\left( 1 \right)'} \left( {K_1 a} \right)}}{{H_m^{\left( 1 \right)} \left( {K_1 a} \right)}} + \frac{{K_1 }}{{K_3 }}\frac{{J'_m \left( {K_c a} \right)}}{{J_m \left( {K_c a} \right)}}} \right]}  \\
   {iK_1 H_m^{\left( 1 \right)} \left( {K_1 a} \right)\left[ {\varepsilon _1 \frac{{H_m^{\left( 1 \right)'} \left( {K_1 a} \right)}}{{H_m^{\left( 1 \right)} \left( {K_1 a} \right)}} - \varepsilon _c \frac{{K_1 }}{{K_c }}\frac{{J'_m \left( {K_c a} \right)}}{{J_m \left( {K_c a} \right)}}} \right]} & {K_x m\frac{{H_m^{\left( 1 \right)} \left( {K_1 a} \right)}}{{k_0 a}}\left( {\frac{{K_1^2 }}{{K_c^2 }} - 1} \right)}  \\
\end{array}} \right)^{ - 1}  \cdot \nonumber \\ 
  &\cdot& \left( {\begin{array}{@{\mkern0mu} cc @{\mkern0mu}}
   {K_x m\frac{{J_m \left( {K_1 a} \right)}}{{k_0 a}}\left( {\frac{{K_1^2 }}{{K_c^2 }} - 1} \right)} & {iK_1 J_m \left( {K_1 a} \right)\left[ { - \frac{{J'_m \left( {K_1 a} \right)}}{{J_m \left( {K_1 a} \right)}} + \frac{{K_1 }}{{K_3 }}\frac{{J'_m \left( {K_c a} \right)}}{{J_m \left( {K_{3} a} \right)}}} \right]}  \\
   {iK_1 J_m \left( {K_1 a} \right)\left[ {\varepsilon _1 \frac{{J'_m \left( {K_1 a} \right)}}{{J_m \left( {K_1 a} \right)}} - \varepsilon _c \frac{{K_1 }}{{K_3 }}\frac{{J'_m \left( {K_c a} \right)}}{{J_m \left( {K_c a} \right)}}} \right]} & {K_x m\frac{{J_m \left( {K_1 a} \right)}}{{k_0 a}}\left( {\frac{{K_1^2 }}{{K_c^2 }} - 1} \right)}  \\
\end{array}} \right).
\end{eqnarray}
After self-consistently assuming that only the $(2M+1)$  cylindrical harmonics with $m = -M,...,M$ are relavant, Eq.(\ref{system}) can be casted as the linear system of algeraic equations
\begin{equation}
\def\arraystretch{1.3}
\sum\limits_{n =  - M}^M {\left[ {\delta _{nm} M^{ - 1} \left( m \right) + \left( { - 1} \right)^{n + 1} U\left( {m + n} \right)} \right]\left( {\begin{array}{@{\mkern0mu} c @{\mkern0mu}}
   {p^{\rm TM} \left( n \right)}  \\
   {p^{\rm TE} \left( n \right)}  \\
\end{array}} \right)}  = \left( {\begin{array}{@{\mkern0mu} c @{\mkern0mu}}
   {Q^{\rm TM} \left( m \right)}  \\
   {Q^{\rm TE} \left( m \right)}  \\
\end{array}} \right),
\end{equation}
which, once solved, provides the coefficients ${p^{\rm TM} \left( m \right), p^{\rm TE} \left( m \right)}$  and hence the fields $F^{(\rm ci)}$ (see the first of Eqs.(\ref{Fci})) and $F^{(\rm cr)}$, $F^{(\rm ct)}$ (see Eqs.(\ref{Uci}) and (\ref{FcrFct})). In addition, the electromagnetic matching conditions at $\rho = a$ also yield
\begin{equation}
\def\arraystretch{1.3}
\left( {\begin{array}{@{\mkern0mu} c @{\mkern0mu}}
   {q^{\rm TM} \left( m \right)}  \\
   {q^{\rm TE} \left( m \right)}  \\
\end{array}} \right) = \frac{{K_1^2 }}{{K_c^2 J_m \left( {K_c a} \right)}}\left[ {J_m \left( {K_1 a} \right)M^{ - 1} \left( n \right) + H_m^{\left( 1 \right)} \left( {K_1 a} \right)} \right]\left( {\begin{array}{@{\mkern0mu} c @{\mkern0mu}}
   {p^{\rm TM} \left( m \right)}  \\
   {p^{\rm TE} \left( m \right)}  \\
\end{array}} \right)
\end{equation}
allowing the evaluation of the field $F^{(\rm ci)}$ inside the cylinder (see the second of Eqs.(\ref{Fci})) once ${p^{\rm TM} \left( m \right), p^{\rm TE} \left( m \right)}$ are known.

\section{Evaluation of the scattering differential cross length}
Once the scattering problem is solved through the machinery developed in Sec.S3, it is possible to evaluate the radiation scattered by the nanowire  into the superstrate from the far field behavior of the field $F^{(\rm ct)}$ of Eqs.(\ref{FcrFct}). Since the mirror symmetric plane waves select the photon wavevector component $k_x = K_x = k_0 \sqrt {\varepsilon _1 } \sin \Theta$ (delta function $\delta(k_x - K_x)$ in Eq.(\ref{Uci})), only the transmitted photons of $F^{(\rm ct)}$  in the parallel momentum range $k_y^2  < {k_0^2 \varepsilon _2  - K_x^2 }$ survive in the far field so that radiation is scattered in the superstrate only if $K_2  = \sqrt {k_0^2 \varepsilon _2  - K_x^2 }$ is positive real. Accordingly we will restrict the analysis to mirror symmetric plane waves such that
\begin{equation}
\Theta  < \arcsin \sqrt {\frac{{\varepsilon _2 }}{{\varepsilon _1 }}} 
\end{equation}
in order to avoid total internal reflection of all the photons generated by the cylinder in the substrate. After introducing cylindrical coordinates $y = r\cos \psi$, $z = r\sin \psi$ and using the relation 
\begin{equation}
\int\limits_{ - \infty }^{ - \infty } {dk_y } e^{ik_y y} e^{ik_{2z} z} F\left( {k_y } \right) \cong e^{i\left( {K_2 r - \frac{\pi }{4}} \right)} \sqrt {\frac{{2\pi }}{{K_2 r}}} \left[ {k_{2z} F\left( {k_y } \right)} \right]_{\scriptstyle k_y  = K_2 \cos \psi  \hfill \atop 
  \scriptstyle k_{2z}  = K_2 \sin \psi  \hfill} 
\end{equation}
providing the leading order of the asymptotic expansion of the Fourier integral (for $k_0 r \rightarrow \infty$ and $\sin \psi >0$), from the second of Eqs.(\ref{FcrFct}) we get 

\begin{equation} \label{FarField}
\def\arraystretch{1.3}
F^{\left( {\rm ct} \right)}  = \sqrt {\frac{{2\pi }}{{K_2 r}}} e^{i\left( {K_2 r - \frac{\pi }{4} + K_x x} \right)} \left[ {\left( {\begin{array}{@{\mkern0mu} cc @{\mkern0mu}}
   {K_2 \sin \psi {\bf{u}}_{\rm S} } & {\left( {K_2 \sin \psi {\bf{u}}_{\rm P}  - k_\parallel  {\bf{e}}_z } \right)}  \\
   { - \frac{{K_2 \sin \psi }}{{k_0 }}\left( {K_2 \sin \psi {\bf{u}}_{\rm P}  - k_\parallel  {\bf{e}}_z } \right)} & {k_0 \varepsilon _2 {\bf{u}}_{\rm S} }  \\
\end{array}} \right)\left( {\begin{array}{@{\mkern0mu} c @{\mkern0mu}}
   {\tilde U_{\rm S}^{\left( {\rm ct} \right)} }  \\
   {\tilde U_{\rm P}^{\left( {\rm ct} \right)} }  \\
\end{array}} \right)} \right]_{{\bf{k}}_\parallel   = K_x {\bf{e}}_x  + K_2 \cos \psi {\bf{e}}_y },
\end{equation}
where
\begin{equation}
\def\arraystretch{1.3}
\left( {\begin{array}{@{\mkern0mu} c @{\mkern0mu}}
   {\tilde U_{\rm S}^{\left( {\rm ct} \right)} }  \\
   {\tilde U_{\rm P}^{\left( {\rm ct} \right)} }  \\
\end{array}} \right) = \left( {\begin{array}{@{\mkern0mu} cc @{\mkern0mu}}
   {T_{\rm SS} } & {nT_{\rm SP} }  \\
   {nT_{\rm PS} } & {T_{\rm PP} }  \\
\end{array}} \right)\left[ {\frac{{e^{ik_{1z} a} }}{{\pi k_\parallel  }}\left( {\begin{array}{@{\mkern0mu} cc @{\mkern0mu}}
   { -  \frac{\varepsilon _1 {k_y }}{{k_{1z} }}} & { - \frac{{k_x }}{{k_0 }}}  \\
   {\frac{{k_x k_{1z} }}{{k_0^2 }}} & { - \frac{{k_y }}{{k_0 }}}  \\
\end{array}} \right)\sum\limits_{m =  - M}^M {\left( { - \frac{{ik_y  - k_{1z} }}{{K_1 }}} \right)^m \left( {\begin{array}{@{\mkern0mu} c @{\mkern0mu}}
   {p^{\rm TM} \left( m \right)}  \\
   {p^{\rm TE} \left( m \right)}  \\
\end{array}} \right)} } \right].
\end{equation}
Equation (\ref{FarField}) displays the chararacteristic outgoing cylindrical wave behavior of the far field, as expected, and it can be readly evaluated once the coefficients ${p^{\rm TM} \left( m \right), p^{\rm TE} \left( m \right)}$ are known. The scattered power  
per unit of angle $\psi$ and unit of length $x$ is 
\begin{equation} \label{dP}
\frac{{dP}}{{d\psi dx}} = r{\bf{e}}_r  \cdot {\bf{S}}^{\left( {\rm ct} \right)} 
\end{equation}
where, ${\bf{e}}_r$ is the cylindrical radial unit vector and ${\bf{S}}^{\left( {\rm ct} \right)}  = \frac{1}{2}{\mathop{\rm Re}\nolimits} \left[ {{\bf{E}}^{\left( {\rm ct} \right)*}  \times {\bf{H}}^{\left( {\rm ct} \right)} } \right]$ is the Poyting vector of the field $F^{(\rm ct)}$. Dividing Eq.(\ref{dP}) by the intensity of one of the mirror symmetric plane waves $I^{\left( {\rm ei} \right)}  = \frac{{\sqrt {\varepsilon _1 } }}{{2Z_0 }}\left| {E_0 } \right|^2$, after a straightforard evaluation we get the differential cross length
\begin{equation}
\Lambda \left( \psi  \right) = \frac{1}{{I^{\left( {\rm ei} \right)} }}\frac{{dP}}{{d\psi dx}} = \frac{{2\pi }}{{k_0 \sqrt {\varepsilon _1 } }}\left[ {\left| {\frac{{K_2 \sin \psi }}{{E_0 }}\tilde U_{\rm S}^{\left( {\rm ct} \right)} } \right|^2  + \left| {\frac{{k_0 \sqrt {\varepsilon _2 } }}{{E_0 }}\tilde U_{\rm P}^{\left( {\rm ct} \right)} } \right|^2 } \right]_{{\bf{k}}_\parallel   = K_x {\bf{e}}_x  + K_2 \cos \psi {\bf{e}}_y } 
\end{equation}
which fully characterizes the scattered radiation pattern.

Due to the mirror symmetry breaking produced by the slab chirality, the scattered radiation pattern is not symmetric and its asymmetry can be ascertained by comparing the "right" and "left" cross lengths, ${\int\limits_0^{\pi /2} {d\psi \Lambda } }$ and ${\int\limits_{\pi /2}^\pi {d\psi \Lambda } }$,  associated to the powers scattered by the cylinder into the right and left portions of the superstrate, respectively. Thus we are led to define dissymmetry factor 
\begin{equation}
\Delta  = 2\frac{{\displaystyle \int\limits_0^{\pi /2} {d\psi \Lambda }  - \int\limits_{\pi /2}^\pi  {d\psi \Lambda } }}{{\displaystyle \int\limits_0^{\pi /2} {d\psi \Lambda }  + \int\limits_{\pi /2}^\pi  {d\psi \Lambda } }}
\end{equation}
which quantitatively measures the asymmetry of the radiation pattern.